\documentclass[orivec]{llncs}

\usepackage[T1]{fontenc}

\usepackage[utf8]{inputenc}

\usepackage{graphicx}
\usepackage{makeidx}

\usepackage{url}
\usepackage{svg}

\usepackage{listings}
\usepackage{color}

\definecolor{dkgreen}{rgb}{0,0.6,0}
\definecolor{gray}{rgb}{0.5,0.5,0.5}
\definecolor{mauve}{rgb}{0.58,0,0.82}
\usepackage{newtxtext,newtxmath}

\urldef\springerdoi\url{https://doi.org/10.1007/978-3-319-78024-5_35}

\renewcommand{\thefootnote}{\fnsymbol{footnote}}

\newcommand\blfootnote[1]{%
  \begingroup
  \renewcommand\thefootnote{}\footnote{#1}%
  \addtocounter{footnote}{-1}%
  \endgroup
}

\begin{document}
\frontmatter
\pagestyle{headings}
\newcommand{\fixme}[1]{{\em\bf{[FIXME: #1]}}}

\mainmatter

\title{Dfuntest: A Testing Framework \\for Distributed Applications}

\titlerunning{dfuntest}
\author{Grzegorz Milka \and Krzysztof Rzadca}
\authorrunning{Milka & Rzadca}
\tocauthor{Grzegorz Milka, Krzysztof Rzadca}
\institute{Institute of Informatics, University of Warsaw, Poland\\
  \email{grzegorzmilka@gmail.com}, \email{krz@mimuw.edu.pl}}

\maketitle

\begin{abstract}
New ideas in distributed systems (algorithms or protocols) are commonly tested by simulation, because experimenting with a prototype deployed on a realistic platform is cumbersome.
However, a prototype not only measures performance but also verifies assumptions about the underlying system.
We developed dfuntest --- a testing framework for distributed applications that defines abstractions and test structure, and automates experiments on distributed platforms.
Dfuntest aims to be jUnit's analogue for distributed applications; a framework that enables the programmer to write robust and flexible scenarios of experiments.
Dfuntest requires minimal bindings that specify how to deploy and interact with the application.
Dfuntest's abstractions allow execution of a scenario on a single machine, a cluster, a cloud, or any other distributed infrastructure, e.g. on PlanetLab.
A scenario is a procedure; thus, our framework can be used both for functional tests and for performance measurements.
We show how to use dfuntest to deploy our DHT prototype on 60 PlanetLab nodes and verify whether the prototype maintains a correct topology.
\end{abstract}

\section{Introduction}
While distributed algorithms are usually tested by simulation or
verified formally, many additional insights can be learned from
experimenting with a prototype implementation deployed on a realistic
platform. \blfootnote{Author's version of a paper published in PPAM 2017, Part I, LNCS 10777 Proceedings, Springer. 
  The final publication is available at Springer via
  \springerdoi
}

For instance, research on contract-based p2p storage~\cite{Rzadca2010ReplicaPlacementin,pamies2011enforcing} focused on choosing replication partners; but only when implementing a prototype~\cite{Skowron2013p2pbackup} we realized that there was no protocol specifying how to make such a contract in a dynamically-changing environment.

Unfortunately, performing and reproducing experiments with prototypes is tedious.
Not only a researcher must become a developer (or hire a student) to code---they also need to configure the experimental platform (i.e., the system running the prototype), deploy the prototype and the test data; and then, after completing a test scenario, gather the results and analyze them.
Developers commonly write ad-hoc scripts to automate many of these tasks---these scripts are often repetitive, full of boilerplate, and do not add any new core functionality.
Moreover, such scripts are hard to maintain and port between user
credentials, or experimental platforms (which can be very diverse, from processes running on a single physical machine, to virtual machines rented from a public clouds or machines from research systems such as PlanetLab~\cite{chun2003planetlab}).

In this paper, we describe dfuntest, a testing framework for distributed applications.
Dfuntest's goal is to be jUnit's analogue for distributed applications.
Dfuntest defines a flexible testing pattern that helps to structure tests.
The principal feature of our testing pattern is clear separation of concerns. The framework distinguishes between features common across all distributed applications (such as deploying files to a remote host; or launching a process); common between tests of a single application (e.g., specifying which files should be deployed; or how to launch an application); and specific to a particular test scenario.
Dfuntest describes these abstractions through interfaces and abstract classes, which guide the developer when preparing tests of a specific application.
Moreover, to facilitate  the testing process, dfuntest provides reference implementations for typical experimental platforms: a set of hosts reachable by SSH and processes executing on a local operating system.


The paper is organized as follows. We define a design pattern for testing distributed applications in Section~\ref{sec:abstractions}.
We then describe a concrete implementation of the proposed pattern, the dfuntest framework.
Section~\ref{sec:dfun-architecture} presents its architecture; and Section~\ref{sec:dfunt-impl} the key elements of implementation.
We evaluate dfuntest by testing Ghoul, our distributed hash table (DHT) implementation (Section~\ref{sec:exampl-test-kademl}): we show how to verify that Ghoul maintains a correct DHT topology.

Dfuntest is available with an Open Source license at \url{https://github.com/gregorias/dfuntest}.

\section{Dfuntest Design}\label{sec:dfuntest-design}
In this section we describe the design of our testing framework. To facilitate presentation, we start with an abstraction of a testing process (Section~\ref{sec:abstractions}). We then proceed with the description of the architecture (Section~\ref{sec:dfun-architecture}); and provide the key details of the implementation of this architecture (Section~\ref{sec:dfunt-impl}).

We will use the following vocabulary when describing the architecture.
The \emph{tester} is the framework's user.
The application/system tested by the framework (called the SUT in~\cite{ulrich1999test}) is the \emph{(tested) application}.
This application's deployment consists of many \emph{instances}, which communicate through network.
An instance runs in an \emph{environment}---a remote host, a virtual machine, or a separate directory of a machine.

\subsection{Abstracting a Distributed Application's Testing Process}\label{sec:abstractions}
The structural abstractions of the testing process are a key feature of testing frameworks.
These abstractions are a result of a delicate equilibrium.
They must be general enough not to limit tester's expressiveness; but they must
be feasible to automate by the library code.
We recognize the following phases when running a single test of a distributed
application:
\begin{enumerate}
\item \textbf{Test configuration} The tester decides which scenarios to run, testing environments and parameters.
\item \textbf{Environment preparation} The framework deploys the application and the test data on the target environments..
\item \textbf{Testing} The framework executes the tested application according to a pre-defined \emph{scenario}; the test checks assertions about the application's state.
\item \textbf{Report generation} A report includes the test result and supplementary information for debugging, such as logs.
\item \textbf{Clean up} The framework deletes generated artifacts from remote hosts.
\end{enumerate}

\subsection{Architecture of Dfuntest}\label{sec:dfun-architecture}

\begin{figure}[tb]
  \centering
  \includegraphics[width=\textwidth]{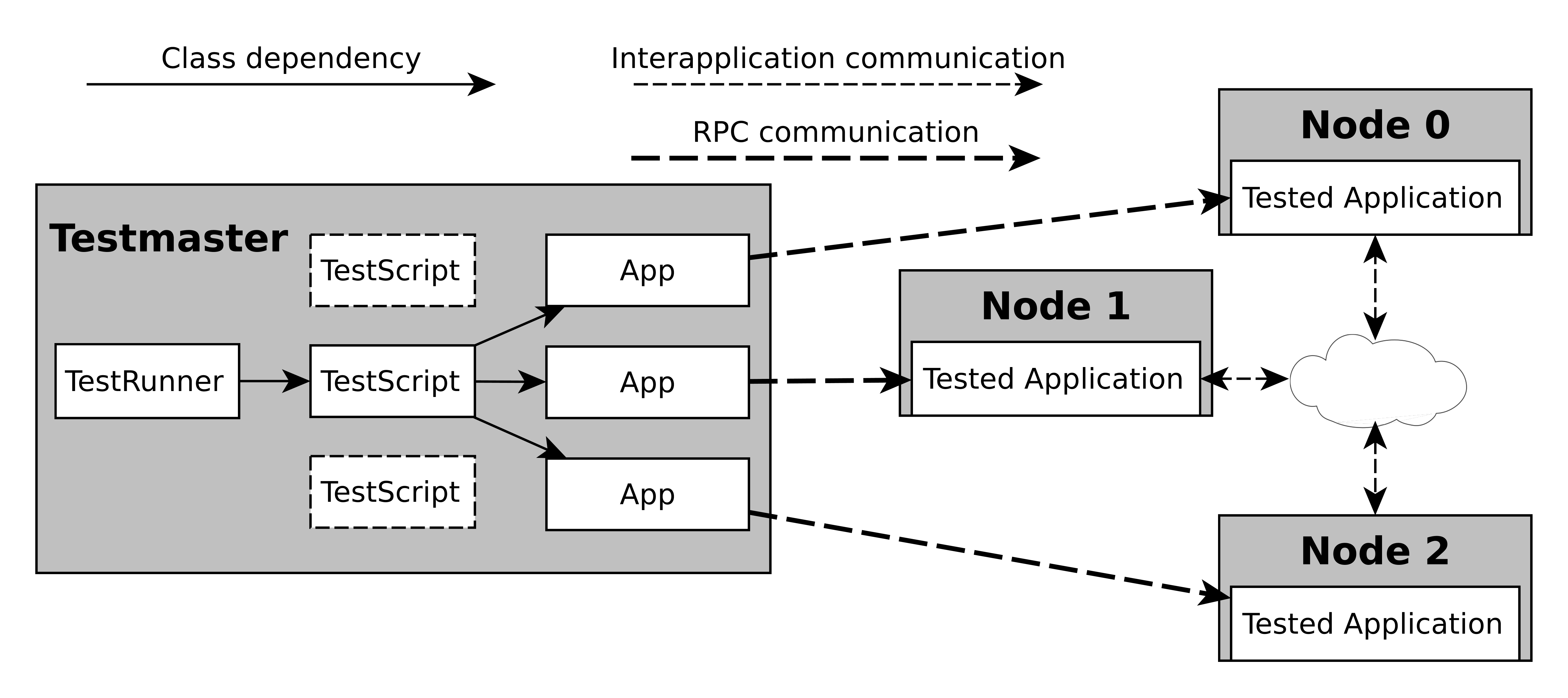}
\caption{Dfuntest during the testing phase. The tested application consists of 3 instances deployed on 3 nodes.
  The dfuntest code runs on the testmaster. The application runs according to a scenario specified in a TestScript. The interactions are instrumented by executing local methods of App objects, which dfuntest then translates to calls (RPC) to interfaces exposed by remote instances of the tested application.}\label{fig:dfuntest-deployment}
\end{figure}

Dfuntest has a centralized architecture: a single entity, the \emph{testmaster}, instruments all phases of the test (yet each phase is inherently distributed; in Section~\ref{sec:rel-work} we further discuss an alternative, distributed design).
Tests scenarios can be fully automated, or highly interactive. They are limited only by the tested application's interface.

In our description, we use standard design patterns~\cite{GoF94DesignPatterns}: a factory and a proxy. Moreover, since the complete architecture is complex, rather than discussing a UML class diagram, we show the key elements in Figure~\ref{fig:dfuntest-deployment}, and then discuss their interactions during a single test. 
However, to motivate certain design decisions, we do not follow the test phases in a sequential order. We start our description from the \emph{testing} phase.

During the testing phase (phase 3 in Section~\ref{sec:abstractions}), the \emph{testmaster} instruments \emph{instances} of the tested application to execute some actions according to a \emph{scenario} (this phase is depicted on Figure~\ref{fig:dfuntest-deployment}).
The scenario, defined by the \emph{tester}, is represented by an object inheriting from the \texttt{TestScript} interface.
the \texttt{TestScript}
declares a single method \texttt{run}; the method executes the testing scenario, checks
assertions, and returns a report.
In the \emph{testmaster}, instances of the tested application are represented by objects of a class inheriting from an \texttt{App}.
An \texttt{App} translates local Java method calls into remote procedure calls (RPC) to a particular instance of the tested application (an \texttt{App} is thus a proxy to a instance of a tested application).
Thus, the tester writes a scenario as she would wrote a standard unit test (calling methods of objects and verifying assertions).The complexity of a distributed system is hidden behind the \texttt{App} (the tester has to implement a specific \texttt{App} subclass for the tested application, but this implementation is reused across many scenarios).


To create the environment for the application on a remote host (phase 2), dfuntest uses standard OS tools to copy, upload, or download files, traverse directories, run processes
etc.  Dfuntest abstracts from concrete implementations of these tools through a proxy
class \texttt{Environment}.
This separation of an \texttt{Environment} from an \texttt{App} gives greater
flexibility and allows to test an application in diverse environments (by using different implementations of the \texttt{Environment}).
One use-case is a test failing on a remote environment: by changing
the remote environment to a local one, 
debugging is faster and easier.

The creation of a remote environment depends strongly on the distributed application (e.g., choosing which test data to put on which hosts). The tester configures the test (phase 1 in Section~\ref{sec:abstractions}) by specifying a script describing how to create a configured environment in a class implementing \texttt{EnvironmentPreparator} interface.
Tester's script uses methods of an \texttt{Environment}: e.g., if an instance needs a particular data file, the script will invoke \texttt{Environment}'s \texttt{copyFilesFromLocalDisk}.

After a test finishes (phase 4), the \texttt{TestScript} is responsible for generating a report documenting its execution. Finally (phase 5), the \texttt{EnvironmentPreparator} cleans up environments and downloads test's artifacts (such as logs produced by instances).

\subsection{Dfuntest Implementation}\label{sec:dfunt-impl}
In this section, we describe how dfuntest maps the abstractions sketched in the previous section to code.
Dfuntest defines a number of interfaces and provides reusable tools (such as concrete \texttt{Environment}s used for interacting with remote hosts) to stitch a coherent testing framework.

\paragraph{\texttt{Environment}} (Figure~\ref{fig:env_interface}) is a proxy
for creating processes and managing files. The goal of this interface is to permit tests to execute in diverse deployment scenarios.
Dfuntest provides two implementations of an \texttt{Environment}:
a \texttt{LocalEnvironment} and an \texttt{SSHEnvironment}.
In a local test, an application is deployed to multiple directories of the testmaster; the \texttt{LocalEnvironment} acts on the provided directory.
An \texttt{SSHEnvironment} connects to a remote host and translates method calls to SSH functions, e.g. \texttt{copy} to \texttt{scp}.
(We use basic ssh as we find it reliable in PlanetLab; however, on more stable platforms further implementations of the \texttt{Environment} can use deployment tools such as Chef~\cite{chefwww}).
A tester may want to add new functions to the \texttt{Environment}. For this reason, we defined other dfuntest interfaces as generics, taking a subclass of the \texttt{Environment} as a parameter.

\begin{figure}[tb]
\begin{lstlisting}
public interface Environment {
  void copyFilesFromLocalDisk(Path srcPath, String destRelPath) throws IOException;
  void copyFilesToLocalDisk(String srcRelPath, Path destPath) throws IOException;
  RemoteProcess runCommand(List<String> command) throws InterruptedException, IOException;
  void removeFile(String relPath) throws IOException;
  ...
}
\end{lstlisting}
\caption{A fragment of the \texttt{Environment} interface}
\vspace{2mm}
\label{fig:env_interface}
\end{figure}

\paragraph{\texttt{EnvironmentPreparator}} (Figure \ref{fig:envprepint}) defines the dependencies between the application and its environment. Using \texttt{Environment}'s methods, the \texttt{EnvironmentPreparator} prepares the environment, collects the test's output, and cleans the environment. As these functions are specific to a distributed application, the tester implements the preparator as a class implementing the \texttt{EnvironmentPreparator} interface.

Some applications depend on many external libraries, or datasets; copying these
files to remote hosts takes time.
To speed-up environment preparation for subsequent tests in a single test suite, we
split the preparation process into two methods:
\texttt{prepare} assumes an empty environment, and thus copies all dependencies;
while \texttt{restore} assumes that all read-only files (libraries or datasets) have been loaded.

\begin{figure}[tb]
\begin{lstlisting}
public interface EnvironmentPreparator<EnvT extends Environment> {
  void prepare(Collection<EnvT> envs) throws IOException;
  void restore(Collection<EnvT> envs) throws IOException;
  void collectOutput(Collection<EnvT> envs, Path destPath);
  void cleanOutput(Collection<EnvT> envs);
  void cleanAll(Collection<EnvT> envs);
}
\end{lstlisting}
\caption{The \texttt{EnvironmentPreparator} interface}
\label{fig:envprepint}
\end{figure}



\vspace{-1em}
\paragraph{\texttt{EnvironmentFactory}, \texttt{AppFactory}} Since
the \texttt{Environment} and the \texttt{App} are meant to be subclassed, dfuntest uses
the factory pattern to hide specific implementation from classes that do not require
it, like a \texttt{TestRunner}.

\vspace{-1em}
\paragraph{\texttt{TestRunner}}
A runner gathers all the classes described above (including a collection of \texttt{TestScript}s to run) and runs the entire testing pipeline. It is a dfuntest equivalent of a
\texttt{Runner} in jUnit.
The \texttt{TestRunner} uses the \texttt{EnvironmentFactory} to
create and prepare remote environments. Then, for each test, it uses the
the \texttt{AppFactory} to create instances of \texttt{App}s (which in turn start
the remote instances of application). Once \texttt{App}s are created,
the \texttt{TestRunner} runs \texttt{TestScript}s, collecting logs and cleaning
environments in-between. Finally it produces a report directory.

\section{Example: Testing Ghoul, a Kademlia DHT Implementation}\label{sec:exampl-test-kademl}
In this section we show what actions are needed to use dfuntest to test a concrete distributed application. In short, a tester first needs to implement an \texttt{App} to translate application's external interface to Java methods; an \texttt{AppFactory} that constructs the newly-created \texttt{App}; and an \texttt{EnvironmentPreparator} to describe how to deploy an instance of the application. Then, a tester implements test scenarios through \texttt{TestScript}s.

A distributed hash table (DHT) is a distributed data structure storing objects indexed by keys (usually large integers) on a possibly large number of machines.
Our distributed file storage system, nebulostore~\cite{nebulowww}, uses a DHT as an index of the stored files.
However, we could not find an open-source DHT implementation that would survive 30 minutes on 50 PlanetLab nodes. 
We thus decided to implement a DHT from scratch.
Our implementation, Ghoul, is based on the Kademlia protocol~\cite{maymounkov2002kademlia}.
Ghoul extends Kademlia with new cryptography functionality, but the tests described below concern the fundamental DHT functions.
Due to space constraints, we show only a single test, and only its most interesting parts; the whole code is available at \url{https://github.com/gregorias/ghoul}.
We run Ghoul's test scenarios on over 60 PlanetLab nodes (so far, scaling is limited by the availability of PlanetLab nodes, rather than dfuntest constraints).

Ghould exposes an external API over HTTP (note that this interface is additional to the basic DHT protocol implementation, which is over UDP).
This external API allows to run typical DHT operations (put/get) and, for testing purposes, to access DHT routing tables of the instance.


\subsection{Preparation of the App and the Environment}
The following code does not need to be changed as long as
the Ghoul API and its requirements remain the same.

\vspace{-1em}\paragraph{\texttt{App}}
\texttt{GhoulApp} extends the proxy \texttt{App} with Ghoul's interface methods.
The \texttt{GhoulApp}'s main responsibility is to translate Java method calls into the RPC-over-HTTP interface of a Ghoul instance.
An example code of those methods is shown in Figure~\ref{fig:app_example}.
The \texttt{GhoulApp} takes an \texttt{Environment} as a parameter (the base class, as to deploy and run Ghoul needs just the standard operations).
To construct a representation of a Ghoul instance, the \texttt{GhoulApp} takes the URI address of the instance's external API; and an \texttt{Environment} object representing an environment on which the instance is deployed.

\begin{figure}[tb]
\begin{lstlisting}
public synchronized void startUp() throws IOException {
  List<String> runCommand = Arrays.asList({mJavaCommand, "-cp", "lib/*:Ghoul.jar", "me.gregorias.Ghoul.interface.Main", "Ghoul.xml"})
  mProcess = mGhoulEnv.runCommandAsynchronously(runCommand);
}
public Collection<NodeInfo> findNodes(Key key) throws IOException {
  WebTarget target = ClientBuild.newClient().target(mUri)
      .path("find_nodes/" + key.toInt());
  NodeInfoCollectionBean beanColl = target.request(
      MediaType.APPLICATION_JSON_TYPE).get(NodeInfoCollectionBean.class);
  return Arrays.asList(beanColl.getNodeInfo()).stream()
      .map(NodeInfoBean::toNodeInfo).collect(Collectors.toList());
}
\end{lstlisting}
\caption{\texttt{GhoulApp}: start a remote Ghoul instance; and use
HTTP-RPC to find neighboring nodes of the instance.}
\label{fig:app_example}
\end{figure}

\vspace{-1em}\paragraph{\texttt{GhoulEnvironmentPreparator}}
copies Ghoul's dependency jar files and configuration files to the target environment using the standard \texttt{Environment} methods (e.g.: \texttt{copyFilesFromLocalDisk}).


\vspace{-1em}\paragraph{\texttt{GhoulAppFactory}} instantiates \texttt{GhoulApp}s given prepared \texttt{Environments}.

\subsection{Test Scenario: Analysis of DHT Routing Tables}
As an example scenario, we test whether the graph induced by Ghoul instances' routing tables is connected (as each DHT node must be able to access every other node).
Figure~\ref{fig:core} shows the \texttt{ConsistencyTestScript} implementing the \texttt{TestScript}.
The script takes as an argument a collection of Apps representing instances on prepared environments. The script starts the instances (as processes) and then orders them to start the DHT routing protocol (for technical reasons Ghoul does not immediately start the protocol).
Then, the script periodically (using \texttt{scheduleCheckerToRunPeriodically}) runs consistency checks implemented in a class \texttt{ConsistencyChecker}.
A period check queries instances' routing tables using instance's HTTP-RPC interface and constructs a connection graph (\texttt{getConnectionGraph}). If the graph is not connected, the test fails.

\begin{figure}[tb]
\begin{lstlisting}
public TestResult run(Collection<GhoulApp> apps) {
  try {
    startUpApps(apps); // start instances (e.g. remote processes)
    startGhouls(apps); // order instances to start DHT routing
  } catch (GhoulException | IOException e) {
    shutDownTest(apps);
    return new TestResult(Type.FAILURE, "Could not start Ghouls.", e);
  }
  mResult = new TestResult(Type.SUCCESS, "Topology was consistent");
  scheduleCheckerToRunPeriodically(new ConsistencyChecker(apps))
  waitTillCheckerFinished();
  shutDownTest(apps);
  return mResult;
}
private class ConsistencyChecker {
  public void run() throws IOException {
    Map<Key, Collection<Key>> graph = getConnectionGraph(mApps);
    ConsistencyResult result = checkConsistency(graph);
    if (result.getType() == ConsistencyResult.Type.INCONSISTENT) {
      mResult = new TestResult(Type.FAILURE, "Graph is not consistent.");
      shutDown();
    }
  }
}
\end{lstlisting}
\caption{\texttt{ConsistencyTestScript} periodically checks whether the graph  induced by DHT routing tables is connected.  (fragment)}
\vspace{-1mm}
\label{fig:core}
\end{figure}

\subsection{Running the Test}

We configured Ghoul build system to create a separate jar package executing the dfuntest described above.
The tester executes the package on a testmaster as a standard Java application.
The main method expects an XML configuration file.
This configuration file contains parameters for setting up
environments and controlling test execution, such as host names and user credentials.
The rest is handled by the dfuntest framework which, after completing the tests, produces a human-readable report directory. This report directory contains a summary report and, for each executed TestScript, a sub-directory with results produced by TestScript and logs copied from environments.


To inject a failure, we changed the value of \emph{bucket size}, a Kademlia parameter describing the maximum number of hosts kept in an entry of the routing table~\cite{maymounkov2002kademlia}. We set the bucket size to 1 (usually it is equal to 20).
Such small bucket should disconnect the graph, because DHT node finding messages will have too little diversity. As expected, the \texttt{ConsistencyTestScript} discovered a disconnected graph, producing a summary clearly pointing to a misbehaving node, 7 (note that for a more compact presentation, this test is executed just on 8 instances).
\vspace{-0.5em}
\begin{lstlisting}[language={},frame={}]
[FAILURE] Found inconsistent graph. The connection graph was:
3: [2, 0, 4]
0: [1, 2, 4]
1: [0, 2, 4]
6: [4, 0]
7: [4, 0]
4: [5, 6, 0]
5: [4, 6, 0]
2: [3, 0, 4]
Its strongly connected components are: [7] [3, 2, 0, 1, 6, 4, 5]
\end{lstlisting}
\vspace{-2em}


\section{Discussion and Related work}\label{sec:rel-work}
Although several frameworks for testing distributed applications have been proposed, we found none that has dfuntest scope and that we could use. To summarize, dfuntest's goal is to be jUnit for distributed applications, i.e., to introduce a single new feature---ability to cope with distributed applications---to a well-understood test ecosystem.

An alternative to pre-determined scenarios
is tracing the execution of the application as it is deployed in production (e.g.~\cite{sigelman2010dapper}) and injecting faults (e.g.~\cite{alvaro2016automating}): these are orthogonal to our approach as they are also used in single-host applications.

Dfuntest uses a centralized control for the testing process (also called a global tester~\cite{ulrich1999test}). 
It is easier to verify global properties of the application once
the whole state is represented in a single process.
A scenario with distributed control can be centralized by exposing the requested behavior in an external interface of the tested application. In contrast, decentralization of a centralized test is more difficult.
However, a centralized control is less scalable, and thus it might prevent the framework from effective testing of larger deployments. In our experiments, we found out that it is not the case for mid-size deployments, as dfuntest managed to instrument 60 hosts on PlanetLab network (we were limited by hosts' availability, rather than dfuntest performance).  We refer to~\cite{ulrich1999test,hierons2008effect,hierons2012oracles} for further discussion.

\cite{ulrich1999test}~proposes a decentralized testing architecture and presents a tool for distributed monitoring and assessment of test events. This tool does not facilitate deployment automation.
In~\cite{tsai2003scenario}, a test scenario defined in an XML file uses the tested application's external SOAP interface.
While dfuntest also uses external interface, we envision that this interface is enriched for particular tests (by adding new methods); moreover, scenarios as Java methods enable greater expressiveness.
In~\cite{hughes2004framework}, the code of the tested application is modified using aspects (thus, the framework tests only Java code).
Given examples focus on monitoring rather than testing---tests can verify how many nodes are, e.g., executing a method.
Similarly,~\cite{butnaru2007p2ptester} focuses on performance measurements.
\cite{de2010testing}~uses annotations, which again limits the applicability of the framework to Java applications.
The scenarios are defined in a pseudo-language that, compared to dfuntest, might increase readability, but also reduce expressiveness.
The framework is more distributed compared to dfuntest, as proxy objects (similar to our \texttt{App}) run on remote hosts.
Remote proxies reduce the need for an external interface;
however, dfuntest centralization helps to check assertions on the state of the whole system.
\cite{torens2010remotetest}~focuses on methods of isolating submodules by emulating some of the components---dfuntest tests the whole distributed application.

To our best knowledge, frameworks described above are not publicly available. In addition to described differences, they do not abstract the remote environment (dfuntest's \texttt{Environment} and \texttt{EnvironmentPreparator}), thus they do not facilitate deployment, nor porting tests between user credentials or testing infrastructures.

\cite{duarte2006gridunit,oliveira2013framework} to speed-up the testing process, distributes execution of test suites using grid~\cite{duarte2006gridunit} or IaaS cloud~\cite{oliveira2013framework}; but the application itself is not distributed. 

\cite{rellermeyer2007building} shows an Eclipse plug-in that creates a GUI for deploying and monitoring OSGi distributed applications. Dfuntest's \texttt{Environment} also deploys applications, but without requiring OSGi-compliance (but requiring explicit dependency management in \texttt{App}).

We continue with the available software for distributed testing.
The Software Testing Automation Framework~\cite{staf} is an open source
project that creates cross-platform, distributed software test environments. It
uses services to provide an uniform interface to environment's resources, such
as file system, process management etc. 
STAF is thus analogous to he \texttt{Environment} abstraction layer in dfuntest.

SmartBear TestComplete~\cite{smartbear} allows to define arbitrary environments and run test jobs sequentially.
SmartBear does not provide any particular mechanism for running a testing scenario or generating a testing report.
Additionally the software requires that the environment has the TestComplete software installed and running and the application uses TestComplete bindings.

Robot Framework~\cite{robot} is a generic test automation framework for
acceptance testing and acceptance test-driven development. Users can define
their testing scenarios in a high-level language resembling natural language.
Robot Framework then automates running and generating a testing report.
Robot does not provide any mechanisms for distributed test control and preparation of a flexible distributed environment.

None of those libraries covers the entire scope of dfuntest framework. What
all of them lack is the ability to code complex testing scenarios, which is provided by \texttt{TestScript} and \texttt{App} abstraction layers.

\section{Conclusions and Perspectives}

We present dfuntest's design pattern for writing distributed tests with centralized control.
Dfuntest offers a coherent and expressive abstraction for distributed testing.
This abstraction allows clean automation of the testing process that in turn also gives more flexible control over the real-world testing environment.
Dfuntest is written in Java, but the tested application may be written in any language, since dfuntest use the application's external interface.

Dfuntest's setup, deployment and clean-up abstractions can be also used to automate basic performance evaluation of a distributed application. In this usage, a test-script can, for instance, measure the delay between an action on an instance and the moment its results propagate to other instances.

\noindent \textbf{Acknowledgements} This research has been supported by
a Polish National Science Center grant Sonata (UMO-2012/07/D/ST6/02440).



\bibliographystyle{splncs03}
\bibliography{CP046}

\begin{thebibliography}{10}
\providecommand{\url}[1]{\texttt{#1}}
\providecommand{\urlprefix}{URL }

\bibitem{smartbear}
Automated software testing tools {TestComplete}.
  \url{http://smartbear.com/product/testcomplete/overview/}, accessed: 26 Sept
  2017

\bibitem{chefwww}
Chef: Deploy new code faster and more frequently. \url{http://www.chef.io},
  accessed: 26 Sept 2017

\bibitem{nebulowww}
Nebulostore: a p2p storage system. \url{http://nebulostore.org}, accessed: 26
  Sept 2017

\bibitem{robot}
Robot framework. \url{http://robotframework.org/}, accessed: 26 Sept 2017

\bibitem{staf}
Software testing automation framework {(STAF)}.
  \url{http://staf.sourceforge.net}, accessed: 27 Sept 2017

\bibitem{alvaro2016automating}
Alvaro, P., Andrus, K., Sanden, C., Rosenthal, C., Basiri, A., Hochstein, L.:
  Automating failure testing research at internet scale. In: SoCC, Proc. pp.
  17--28. ACM (2016)

\bibitem{butnaru2007p2ptester}
Butnaru, B., Dragan, F., Gardarin, G., Manolescu, I., Nguyen, B., Pop, R.,
  Preda, N., Yeh, L.: {P2PTester}: a tool for measuring {P2P} platform
  performance. In: ICDE. pp. 1501--1502 (2007)

\bibitem{chun2003planetlab}
Chun, B., Culler, D., Roscoe, T., Bavier, A., Peterson, L., Wawrzoniak, M.,
  Bowman, M.: Planetlab: an overlay testbed for broad-coverage services. ACM
  SIGCOMM Computer Communication Review  33(3),  3--12 (2003)

\bibitem{de2010testing}
De~Almeida, E.C., Suny{\'e}, G., Le~Traon, Y., Valduriez, P.: Testing
  peer-to-peer systems. Empirical Software Engineering  15(4),  346--379 (2010)

\bibitem{duarte2006gridunit}
Duarte, A., Cirne, W., Brasileiro, F., Machado, P.: Gridunit: software testing
  on the grid. In: ICSE. pp. 779--782. ACM (2006)

\bibitem{GoF94DesignPatterns}
Gamma, E., Helm, R., Johnson, R., Vlissides, J.: Design patterns: elements of
  reusable object-oriented software. Addison-Wesley (1994)

\bibitem{hierons2012oracles}
Hierons, R.M.: Oracles for distributed testing. IEEE Trans. on Software
  Engineering  38(3),  629--641 (2012)

\bibitem{hierons2008effect}
Hierons, R.M., Ural, H.: The effect of the distributed test architecture on the
  power of testing. The Computer Journal  51(4),  497--510 (2008)

\bibitem{hughes2004framework}
Hughes, D., Greenwood, P., Coulson, G.: A framework for testing distributed
  systems. In: P2P. pp. 262--263. IEEE (2004)

\bibitem{maymounkov2002kademlia}
Maymounkov, P., Mazieres, D.: Kademlia: A peer-to-peer information system based
  on the xor metric. In: Peer-to-Peer Systems, pp. 53--65. Springer (2002)

\bibitem{oliveira2013framework}
Oliveira, G.S.D., Duarte, A.: A framework for automated software testing on the
  cloud. In: PDCAT. pp. 344--349. IEEE (2013)

\bibitem{pamies2011enforcing}
Pamies-Juarez, L., García-López, P., S{\'a}nchez-Artigas, M.: Enforcing
  fairness in p2p storage systems using asymmetric reciprocal exchanges. In:
  P2P. pp. 122--131. IEEE (2011)

\bibitem{rellermeyer2007building}
Rellermeyer, J.S., Alonso, G., Roscoe, T.: Building, deploying, and monitoring
  distributed applications with eclipse and r-osgi. In: OOPSLA. pp. 50--54. ACM
  (2007)

\bibitem{Rzadca2010ReplicaPlacementin}
Rzadca, K., Datta, A., Buchegger, S.: Replica placement in p2p storage:
  Complexity and game theoretic analyses. In: ICDCS. pp. 599--609. IEEE (2010)

\bibitem{sigelman2010dapper}
Sigelman, B.H., Barroso, L.A., Burrows, M., Stephenson, P., Plakal, M., Beaver,
  D., Jaspan, S., Shanbhag, C.: Dapper, a large-scale distributed systems
  tracing infrastructure. Tech. rep., Google (2010)

\bibitem{Skowron2013p2pbackup}
Skowron, P., Rzadca, K.: Exploring heterogeneity of unreliable machines for p2p
  backup. In: HPCS. pp. 91--98. IEEE (2013)

\bibitem{torens2010remotetest}
Torens, C., Ebrecht, L.: Remotetest: A framework for testing distributed
  systems. In: ICSEA. pp. 441--446. IEEE (2010)

\bibitem{tsai2003scenario}
Tsai, W.T., Yu, L., Saimi, A., Paul, R.: Scenario-based object-oriented test
  frameworks for testing distributed systems. In: FTDCS. pp. 288--294. IEEE
  (2003)

\bibitem{ulrich1999test}
Ulrich, A.W., Zimmerer, P., Chrobok-Diening, G.: Test architectures for testing
  distributed systems. In: QW (1999)

\end{thebibliography}
\end{document}